\documentclass[epj]{svjour}

\usepackage{graphics}

\begin{document}

\title{Dzyaloshinskii-Moriya interactions effects on the entanglement dynamics of a two qubit xxz spin system in non-Markovian environment}

\author{M.Tchoffo, G.C.Fouokeng$^\dagger$, E.Tendong, L.C.Fai}
\institute{Mesoscopic and Multilayer Structures Laboratory, Department of Physics, Faculty of Science, University of Dschang, Cameroon, P.O.Box: 479 Dschang-Cameroon.}

\date{Received: date / Revised version: date}

\abstract{We investigate the entanglement dynamics of a two-qubit Heisenberg XXZ chain with Dzyaloshinskii-Moriya (DM) interactions, interacting with an anisotropic spin bath in thermal equilibrium at temperature $T$, driven by an external magnetic field $\textbf{B}$ along the z-axis. We establish that, for an initially entangled qubit pair, the DM interactions generate entanglement and enhance it in the revival region. At high temperatures and for weak coupling between the two qubits, the DM interactions preserve entanglement. These effects are weakened when the magnetic field $\textbf{B}$ and the Heisenberg coupling are switched on. If the two-qubits are prepared in an initially separable state, the DM interaction instead has a negative effect on their entanglement. As a whole, entanglement can better be preserved in the spin chain even at high temperatures by increasng the external magnetic field $\textbf{B}$ and the Heisenberg couplings, and by tuning the strength of the DM interaction.
\PACS{03.67.-a, 03.65.Ud, 75.30.Ds, 03.65.Yz}}

\maketitle

\section{Introduction}
\label{intro}
Entanglement, the quantum non-local connection \cite{1} has been studied intensely in recent years due to its potential applications in quantum communication and information processing tasks \cite{2,3,4,5,6}. It has been shown also to plays a fundamental role in the quantum phase transitions, which occur in interacting lattice systems at zero temperature \cite{7}. These Potential applications of entanglement have stimulated research on ways to quantify and control it. Different physical systems have been proposed as reliable candidates for the underlying technology of quantum computing and quantum information processing \cite{8,9}.
A key ingredient for all these applications should be able to manipulate coherently entangled state to provide an efficient computational process. Such coherent manipulation of entangled states has been realized in different systems such as isolated trapped ions \cite{10} and superconducting junctions  \cite{11}. The coherent control of a two-electron spin state in a coupled quantum dot has been achieved experimentally \cite{12,13}, in which the coupling mechanism is the Heisenberg exchange interaction between the electron spins.
Heisenberg spin chains are among one of the major quantum systems, which have been proposed for the physical realization of good qubits needed in the implementation of the quantum computer \cite{14,15,16}.  shown in Ref. \cite{17}, the spin systems suffer from decoherence effects due to the influences of the environmental degrees of freedom on the dynamics of the system and to the unavoidable interactions between the qubits and their environment; these cause also the decay of qubit superposition states (entangled states) into a classical state. The phenomenon known as decoherence, can seriously hinder the various quantum information processing tasks.

 In the dynamics of a quantum spin system, the DM interaction \cite{18,19,20,21,22} has been presented as one of the major spin interactions that induced decoherence.  This spin interactions  arises from the consideration of spin orbit coupling effects in Anderson's super-exchange interaction theory. It is an anisotropic anti-symmetric spins interaction  and plays an important role in the entanglement dynamics of spin qubits. More recently its influence on the entanglement of two qubits in various magnetic spin models \cite{23} and on three Qubits entanglement \cite{24,25} have been studied. The XXZ model encompasses the XX model, XY model, the isotropic Ising model and the XXX model which are all relevant for QIP.

 Understanding, quantifying and exploring entanglement dynamics may provide an answer for many questions regarding the decoherence behaviour of quantum spin systems. In the Heinsenberg spin chain, the interaction with the spin bath system often leads to strong non-Markovian behavior. That is, to study the dynamics of such system, the usual Markovian quantum master equations, which are widely used in the area of atomic physics and quantum optics, may fail for many spin bath models. Therefore, it becomes more and more important to develop methods that are capable of going beyond the Markovian approximation.
Recently, the exact dynamics of a two qubit chain in an XY environment have been studied using a simple mathematical technique based on a unitary linear transformation \cite{26} from where the authors have shown that the behavior of the system was extremely non Markovian. In Ref.\cite{27} the study made on the dynamics of a single spin in a spin star environment, using exact methods and various approximation techniques, reveal that the Markovian approximations perform poorly.
In this paper, we study analytically and numerically the exact entanglement dynamics of a two qubit Heisenberg XXZ spin chain with DM interactions interacting with a spin bath in the presence of an external magnetic field using a simple mathematical technique based on a unitary linear transformation \cite{28,29}. We examine the effects of the external magnetic field strength, temperature, intra-bath coupling strength, system bath coupling and anisotropy of the two qubit spin chain, on the entanglement dynamics, considering the interactions of the qubit systems with the environment. We calculate the concurrence of a qubit pair, for an initially disentangled state and for an initially maximally entangled state.

The organization of the paper is as follows: in section 2 we present a brief description of the theoretical approach used and the model for simulating the two qubit XXZ spin chain with DM interactions interacting with a spin bath. In section 3, to study the entanglement dynamics of the model system presented in section 2, we evaluated the concurrence that quantifies the degree of the pair-wise entanglement between the two central qubits and then conclude with discussion of our findings in Section 4.
\section{Theoretical Approach and the Model Hamiltonian}
\label{sec:1}The model used here describes  two coupled spin qubit interacting with a spin bath in the presence of an external magnetic field oriented in the z-direction via XXZ Heisenberg interactions alongside DM interactions, which are considered with both spin chain and spin bath. The environment is modeled here as a one dimensional Heisenberg XY chain with nearest neighbor spin couplings \cite{23}. The total Hamiltonian of the system described above together with the DM interactions can be written in the form

\begin{equation}\label{a1}
  H=H_{S}+H_{SB}+H_{B}
\end{equation}
with\\

$H_{S}=\mu_{0}(S_{01}^{z}+S_{02}^{z})+\Omega(S_{01}^{+}S_{02}^{-}$
 \begin{equation}\label{a2}
  +S_{01}^{-}S_{02}^{+})+\Gamma_{z}S_{01}^{z}S_{O2}^{z}+id_{z}(S_{01}^{+}S_{02}^{-}-S_{01}^{-}S_{02}^{+})
\end{equation}
\begin{equation}\label{a3}
H_{SB}=\frac{g}{\sqrt{N}}\{(S_{01}^{+}+S_{02}^{+})\sum_{a=1}^{N}S_{a}^{-}+(S_{01}^{-}+S_{02}^{-})\sum_{a=1}^{N}S_{a}^{+}\}
\end{equation}
$H_{B}= \sum_{a\neq b}^{N}\{\frac{g}{N}(S_{a}^{+}S_{b}^{-}+S_{a}^{-}S_{b}^{+})$
\begin{equation}\label{a4}
+i\frac{D_{z}}{N}(S_{a}^{+}S_{b}^{-}-S_{a}^{-}S_{b}^{+})\}+\sum_{a=1}^{N}\frac{2\gamma}{N}S_{a}^{z}
\end{equation}
Here $\mu_{0}$ represents the strength of the coupling of the two spin qubits to the external magnetic field, $\Omega$ is the coupling strength between the two spin qubits while $\Gamma_{z}$  represents the coupling strength in the z-direction of the XXZ chain. $S_{0i}^{\pm}(i=1,2)$, represent the spin creation and annihilation operators for the two qubit spin chains while $S_{i}^{\pm}(i=a,b)$ represent the spin creation and annihilation operators for the bath spins. $\gamma$, represents the strength of coupling of the bath spins with the external magnetic field, $g_{0}$ and  $g$ are respectively the spin system-bath coupling strength and the intra-bath coupling strength. $D_{z}$ and $d_{z}$  represent the  z-component of the DM coupling vector between the bath spins and between the two spin qubits. Finally $N$ represents the number of spins in the bath. All the coupling strengths are rescaled so that the free energy of the system remains finite when $N\longrightarrow \infty$.

The form of the above Hamiltonian chosen in this work is due to its relevance for various QIP tasks and it models the environment as closely as possible so that the effects of the environment on the dynamics of the central spin can fully be taken into account. A similar Hamiltonian has been examined recently in \cite{22}. By introducing the collective angular momentum operators
\begin{equation}\label{a5}
  \Gamma^{\pm}=\sum_{a=1}^{N} S_{a}^{\pm}     ;      \Gamma^{z}=\sum_{a=1}^{N} S_{a}^{z}
\end{equation}
we rewrite the Hamiltonians (3) and (4) as
\begin{equation}\label{a6}
  H_{SB}=\frac{g}{\sqrt{N}}\{(S_{01}^{+}+S_{02}^{+})\Gamma^{-}+(S_{01}^{-}+S_{02}^{-})\Gamma^{+}\}
\end{equation}
$ H_{B}=\frac{g}{N}\{(\Gamma^{+}\Gamma^{-}+\Gamma^{-}\Gamma^{+})$
\begin{equation}\label{a7}
 +i\frac{D_{z}}{N}(\Gamma^{+}\Gamma^{-}+\Gamma^{-}\Gamma^{+})\}+2\frac{\gamma}{N}\Gamma^{z}-g-i\frac{D_{z}}{N}\Gamma^{z}
\end{equation}
 							
The low temperature excitation spectrum of the system can be obtained by introducing the following Holstein-Primakoff transformation
\begin{equation}\label{a8}
  \Gamma^{+}=b^{+}\sqrt{(2\vartheta-b^{+}b)};  \Gamma^{-}=\sqrt{(2\vartheta-b^{+}b)}b \Gamma^{z}=\vartheta -b^{+}b
\end{equation}
Where $\vartheta$ denotes the length of the collective environment pseudo-spin $\frac{N}{2}$. Thus;
\begin{equation}\label{a9}
  N=2 \vartheta
\end{equation}
The above transformation transforms the spin operators $\Gamma^{+}$, $\Gamma^{-}$ and $\Gamma^{z}$ into bosonic creation and annihilation operators $b^{+}$ and $b^{-}$ obeying the commutation relation  $[b^{+}b]=1$. Thus, in the thermodynamic limit ($N\longrightarrow\infty$) the Hamiltonians (6) and (7) become
\begin{equation}\label{a10}
  H_{SB}= g_{0}[(S_{01}^{+}+S_{02}^{+})b+(S_{01}^{-}+S_{02}^{-})b^{+}]
\end{equation}
\begin{equation}\label{a11}
  H_{B}= 2gb^{+}b-2iD_{z}+\gamma
\end{equation}
 Equations (2), (10) and (11) are the Hamiltonian of a two coupled spin qubits system interacting with a single-mode thermal bosonic bath with DM interactions both in the bath and in the two qubit chains. We note here that due to the high symmetry of the model, the coupling to the environment is actually represented by a coupling to a single collective environment spin. The effect of this single-mode environment on the dynamics of the two coupled qubits is extremely non-Markovian hence the traditional master equations used in describing the Markovian dynamics of open quantum systems, cannot be used in this case.
		We assume that the initial state of the system-bath is a separable state so its initial density matrix can be written in the form
\begin{equation}\label{a12}
  \rho(0)=|\varphi(0)\rangle \langle\varphi(0)|\otimes\rho_{B}
\end{equation}
The density matrix of the spin bath $\rho_{B}$ satisfies the Boltzmann distribution, i.e;
\begin{equation}\label{a13}
  \rho_{B}=\frac{1}{Z}e^{-\frac{H_{B}}{T}}
\end{equation}
where $Z=Tr(e^{-\frac{H_{B}}{T})}$ is the partition function. Here $Tr$ denotes the trace with $T=K_{B}\tau$ where $\tau $  is the temperature and $K_{B}$  the Boltzmann constant (subsequently we simply refer to $T$ as the temperature). The partition function $Z$ is given by;
\begin{equation}\label{a14}
  Z=e^{2iD_{z}-\gamma}\Big(\frac{1}{1-e^{-\frac{2g}{T}}}\Big)
\end{equation}
At absolute zero temperature, no excitation will exist. The bath is in a thoroughly polarized state with all spins down. The most general form of an initial pure state of the two-qubit system can be written as:
\begin{equation}\label{a15}
   |\varphi(0)\rangle=\alpha|00\rangle + \varepsilon|01\rangle +\delta|10\rangle + \beta|11\rangle
\end{equation}
with the normalization condition   yielding
\begin{equation}\label{a16}
  |\alpha|^{2}+|\varepsilon|^{2}+|\delta|^{2}+|\beta|^{2}=1
\end{equation}
For analytic simplicity, we set $\varepsilon=\delta=0$. So the initial state can be written as
\begin{equation}\label{a17}
   |\varphi(0)\rangle=\alpha|00\rangle + \beta|11\rangle
\end{equation}
and the initial density matrix of the system plus bath takes the form:
\begin{equation}\label{a18}
   \rho(0)=(\alpha|00\rangle + \beta|11\rangle)(\langle 00|\alpha^{*}+\langle11|\beta^{*})\otimes \frac{1}{Z}(e^{-\frac{H_{B}}{T}})
\end{equation}
We note that the time dependent density matrix of the system coupled to the bath obeys to the folowing relation
\begin{equation}\label{a19}
   \rho_{s}(t)=Tr_{B}(\rho(t))
\end{equation}
where $U(t)=e^{iHt}$, is the unitary time evolution operator. The qubit system alone does not evolve in a unitary manner. The dynamics of the qubit system alone is obtain by tracing over  the bath modes in order to obtain the reduced density matrix of the qubit system $\rho_{s}(t)$
\begin{equation}\label{a20}
   \rho(t)=U^{*}(t)\rho(0)U(t)
\end{equation}
 where $Tr_{B}$ denotes the partial trace of the density matrix taken over the bath modes. We obtain the reduced density matrix of the form:\\

$\rho_{s}(t)=Tr_{B}\Big\{(\frac{1}{Z})\Big(|\alpha|^{2}e^{-iHt}|00\rangle e^{-\frac{H_{B}}{T}}\langle 00|e^{iHt}$
\begin{equation}\label{a21}
+\alpha\beta^{*}e^{-iHt}|00\rangle e^{-\frac{H_{B}}{T}}\langle11|e^{iHt}
\end{equation}
$+\alpha^{*}\beta e^{-iHt}|11\rangle e^{-\frac{H_{B}}{T}}\langle00|e^{iHt}+|\beta|^{2}e^{-iHt}|11\rangle e^{-\frac{H_{B}}{T}}\langle 11|e^{iHt} \Big)\Big\}$\\

In order to obtain the full form of the reduced density matrix, we need to evaluate the time evolution of the initial qubit state: $e^{-iHt}|00\rangle$, and  $e^{-iHt}|11\rangle$ . We observe that by applying the time dependent Schrodinger equation,
\begin{equation}\label{a22}
   i\frac{d}{dt}|\varphi(t)\rangle=H|\varphi(t)\rangle
\end{equation}
where
\begin{equation}\label{a23}
|\varphi(t)\rangle=U(t)(\alpha |00\rangle+ \beta|11\rangle)
\end{equation}
From the total Hamiltonian $H$, we can see that it consists of operators of the form $S_0i^+ $ , $S_0i^- $ (i=1,2) which change the state of the $ i^th $ spin from   $|0\rangle$ to $|1\rangle $and from  $|1\rangle $, to  $|0\rangle $ respectively. Thus, the qubit will evolve from the initial pure state into the most general mixed state as follows
\begin{equation}\label{a24}
   e^{-iHt}\mid11\rangle = \Theta \mid00\rangle + \Upsilon \mid01\rangle + \varrho \mid10\rangle+  \flat\mid11\rangle
\end{equation}
\begin{equation}\label{a25}
   e^{-iHt}|00\rangle=\Im|00\rangle + \clubsuit|01\rangle +\S|10\rangle + \pounds|11\rangle
\end{equation}
where $\Theta,\Upsilon,\varrho,\flat,\Im,\clubsuit,\S,\pounds$  are functions of $b^+ $, $b$ and $t$ . Thus:\\

$|\varphi(t)\rangle= \alpha e^{-iHt}|00\rangle + \beta e^{-iHt}|11\rangle = \alpha (\Im|00\rangle + \clubsuit|01\rangle$
\begin{equation}\label{a26}
 +\S|10\rangle + \pounds|11\rangle)+  \beta (\Theta|00\rangle + \Upsilon|01\rangle + \varrho|10\rangle + \flat|11\rangle)
\end{equation}
To obtain the exact form of the reduced density matrix $\rho_{s}(t)$ we need to evaluate the form of the expressions $\Theta,\Upsilon,\varrho,\flat,\Im,\clubsuit,\S,\pounds$. From equation (22) it follows that:
\begin{equation}\label{a27}
   i\frac{d}{dt}(e^{-iHt}|11\rangle)=H (e^{-iHt}|11\rangle)
\end{equation}
and
\begin{equation}\label{a28}
   i\frac{d}{dt}(e^{-iHt}|00\rangle)=H (e^{-iHt}|00\rangle)
\end{equation}
With the initial conditions from being  $\Theta0)=\Upsilon(0)=\flat(0)=0$ and $\flat(0)=1$ , and with the  following set of  transformations
\begin{equation}\label{a29}
  \Theta=-b^{+}b e^{-i(2g(b^{+}b+1)+u)t}\Theta_{1}
\end{equation}
\begin{equation}\label{a30}
  \Upsilon=-b^{+} e^{-i(2g(b^{+}b+1)+u)t}\Upsilon_{1}
\end{equation}
\begin{equation}\label{a31}
  \varrho=-b^{+} e^{-i(2g(b^{+}b+1)+u)t}\varrho_{1}
\end{equation}
\begin{equation}\label{a32}
  \flat= e^{-i(2g(b^{+}b+1)+u)t}\flat_{1}
\end{equation}
the following first order four differential equations are derived
\begin{equation}\label{a33}
   i\frac{d}{dt}\Theta_{1}=i(\mu_{0}-2g)\Theta_{1}-ig_{0}(\Upsilon_{1}+\varrho_{1})
\end{equation}
\begin{equation}\label{a34}
   i\frac{d}{dt}\Upsilon_{1}=2i\Gamma_{z}\Upsilon_{1}-i(\Omega -id_{z})\varrho_{1}-ig_{0}(\hat{n}+2)\Theta_{1}-ig_{0}\flat_{1}
\end{equation}

\begin{equation}\label{a35}
   i\frac{d}{dt}\varrho_{1}=2i\Gamma_{z}\varrho_{1}-i(\Omega -id_{z})\Upsilon_{1}-ig_{0}(\hat{n}+2)\Theta_{1}-ig_{0}\flat_{1}
\end{equation}

\begin{equation}\label{a36}
   i\frac{d}{dt}\flat_{1}=i(\mu_{0}-2g)\flat_{1}-ig_{0}(\hat{n}+1)(\Upsilon_{1}+\varrho_{1})
\end{equation}\\

where $U=\gamma-2iD_{z}+\Gamma_{z}$  and  $b^{+}b=\hat{n}$. The solution of the coupled differential equations (27) is obtained analytically through the initial conditions (29) to (32) for the case $d_{z}=0$ with the resonant condition $\mu_{0}=2g$ ; the external magnetic field can easily be tuned to satisfy this condition. However the numerical results are present for the case where $d_{z}\neq0$. Thus, the following analytical solutions are obtained \\

$\Theta_{1} (t)=  \frac{-1}{3+2\hat{n}}$
\begin{equation}\label{a37}
                                + \frac{2g_o^2}{\sqrt{(2\Gamma_z-\Omega)^2+8g_o^2 (3+2\hat{n})}}\Big \{\frac{e^{i\ell_{1}t}}{\ell_{1}}-\frac{e^{i\ell_{2}t}}{\ell_{2}}\Big \}
\end{equation}

\begin{equation}\label{a38}
   \Upsilon_{1} (t)=\varrho_{1} (t)=- \frac{g_o^2}{\sqrt{(2\Gamma_z-\Omega)^2+8g_o^2 (3+2\hat{n})}}\{e^{i\ell_{1}t}-e^{i\ell_{2}t} \}
\end{equation}. \\

$\flat_{1}(t)= \frac{2+\hat{n}}{3+2\hat{n}}$
	\begin{equation}\label{a39}
                                + \frac{2g_o^2(1+\hat{n})}{\sqrt{(2\Gamma_z-\Omega)^2+8g_o^2 (3+2\hat{n})}}\Big \{\frac{e^{i\ell_{1}t}}{\ell_{1}}-\frac{e^{i\ell_{2}t}}{\ell_{2}}\Big \}
\end{equation}
with
 \begin{equation}\label{a40}
   \ell_{1,2}=\frac{(2\ell_{z}-\Omega)\pm \sqrt{(2\Gamma_{z}-\Omega)^{2}+ 8g_o^2 (3+2\hat{n})}}{2}
\end{equation}
From equation (28), with the initial conditions from being  $\Im(0)=\clubsuit(0)=\S(0)=0$ and $\pounds(0)=1$, we find also with the following transformations
 \begin{equation}\label{a41}
  \Im=e^{-i(2g(b^{+}b-1)+u)t}\Im_{1}
\end{equation}
\begin{equation}\label{a42}
  \clubsuit=b e^{-i(2g(b^{+}b-1)+u)t}\clubsuit_{1}
\end{equation}
\begin{equation}\label{a43}
  \S=b e^{-i(2g(b^{+}b-1)+u)t}\S_{1}
\end{equation}
\begin{equation}\label{a44}
  \pounds= bb^{+}e^{-i(2g(b^{+}b-1)+u)t}\pounds_{1}
\end{equation}
and using the first order differential equations such as in (33) to (36) that the following analytical solutions are obtained:\\

$\Im_{1}(t)=  \frac{\hat{n}-2}{2\hat{n}-1}$
\begin{equation}\label{a45}
 + \frac{2g_o^2\hat{n}}{\sqrt{(2\Gamma_z-\Omega)^2+8g_o^2 (3+2\hat{n})}}\Big \{\frac{e^{i\ell_{1}'t}}{\ell_{1}'}-\frac{e^{i\ell_{2}'t}}{\ell_{2}'}\Big \}
\end{equation}

$\pounds_{1}(t)= \frac{-1}{(2\hat{n}-1)}$
\begin{equation}\label{a46}
 + \frac{2g_o^2\hat{n}}{\sqrt{(2\Gamma_z-\Omega)^2+8g_o^2 (3+2\hat{n})}}\Big \{\frac{e^{i\ell_{1}'t}}{\ell_{1}'}-\frac{e^{i\ell_{2}'t}}{\ell_{2}'}\Big \}
\end{equation}
\begin{equation}\label{a47}
 \clubsuit_{1}(t)= \S_{1}(t)=-\frac{g_o^2}{\sqrt{(2\Gamma_z-\Omega)^2+8g_o^2 (3+2\hat{n})}}\{e^{i\ell_{1}'t}-e^{i\ell_{2}'t} \}
\end{equation}
where
\begin{equation}\label{a48}
   \ell_{1,2}'=\frac{(2\Gamma_{z}-\Omega)\pm \sqrt{(2\Gamma_{z}-\Omega)^{2}+ 8g_o^2 (3+2\hat{n})}}{2}.
\end{equation}
 Thus the exact form of the reduced density matrix is then obtained by tracing over the bath modes and replacing the operator $\hat{n}$ by its Eigen value $n$ as:

\begin{equation}\label{a49}
   \rho_{s}(t)=\left(
                       \begin{array}{cccc}
                         \rho_{11} & 0 & 0 & \rho_{14} \\
                         0 & \rho_{22} & \rho_{23}& 0 \\
                         0&\rho_{32} & \rho_{33}& 0 \\
                         \rho_{14}^{*} & 0 & 0 & \rho_{44}\\
                       \end{array}
                     \right)
                     \end{equation}
where \\

$ \rho_{11}=\big(\frac{1}{Z}\big)\Big(|\alpha|^{2}\sum_{n=0}^{\infty}\Im_{1}\Im_{1}^{*}e^{-\frac{2gn-2iD_{z}+\gamma}{T}}$
\begin{equation}\label{a50}
     +|\beta|^{2}\sum_{n=0}^{\infty}(n+1)(n+2)\Theta_{1}\Theta_{1}^{*}e^{-\frac{2gn-2iD_{z}+\gamma}{T}}\Big)
\end{equation}

\begin{equation}\label{a51}
  \rho_{14}=\big(\frac{1}{Z}\big)\Big(\alpha \beta^{*} \sum_{n=0}^{\infty}\Im_{1}\flat_{1}^{*}e^{-\frac{2gn-2iD_{z}+\gamma}{T}}e^{4igt}
\end{equation}

$\rho_{22}=\rho_{23}=\rho_{32}=$\\

$=\rho_{33}= \big(\frac{1}{Z}\big)\Big(|\alpha|^{2}\sum_{n=0}^{\infty}\clubsuit_{1}\clubsuit_{1}^{*}n e^{-\frac{2gn-2iD_{z}+\gamma}{T}}$
\begin{equation}\label{a52}
     +|\beta|^{2}\sum_{n=0}^{\infty}(n+1)\Upsilon_{1}\Upsilon_{1}^{*}e^{-\frac{2gn-2iD_{z}+\gamma}{T}}\Big)
\end{equation}

$\rho_{44}= \big(\frac{1}{Z}\big)\Big(|\alpha|^{2}\sum_{n=0}^{\infty}\pounds_{1}\pounds_{1}^{*}n(n-1) e^{-\frac{2gn-2iD_{z}+\gamma}{T}}$
\begin{equation}\label{a53}
+|\beta|^{2}\sum_{n=0}^{\infty}(n+1)\flat_{1}\flat_{1}^{*}e^{-\frac{2gn-2iD_{z}+\gamma}{T}}\Big)
\end{equation}
The solutions obtained in equations (50) to (53)  should be used to evaluate the concurrence of the coupled system.
\section{Entanglement dynamics}
\label{sec:2}
For the reduce density matrix given in equation (49) the concurrence quantifying the degree of the pair-wise entanglement between the two central qubits is defined as \cite{30}:
\begin{equation}\label{a54}
  C_{12}=max\{\ell_{1}-\ell_{2}-\ell_{3}-\ell_{4}, 0\}
\end{equation}
where $\ell_{1}$, $\ell_{2}$ , $\ell_{3}$ , $\ell_{4}$  are the squaroots of the Eigen values in order of decreasing magnitude of the operator:
\begin{equation}\label{a55}
 R_{12}=\rho_{s}(\sigma^{y}\otimes\sigma^{y})\rho_{s}^{*}(\sigma^{y}\otimes\sigma^{y})
\end{equation}.
As a measure of the degree of entanglement the concurrence varies from 0 to a maximum value 1. If the concurrence is equal to zero then the two states are said to be completely disentangled or separable while a concurrence $C_{12}= 1 $.  The condition $C_{12}=1$ means that the two states are maximally entangled.
The Eigen values of $R_{12}$ are found to be:

\begin{eqnarray*}
  \ell_{1} &=& \sqrt{\rho_{11}\rho_{44}}+|\rho_{14}| \\
  \ell_{2} &=& \sqrt{\rho_{11}\rho_{44}}-|\rho_{14}|  \\
  \ell_{3} &=& 2\rho_{22} \\
  \ell_{4} &=&  0
\end{eqnarray*}

We shall present our results here for $\Gamma_{z}\geq0$ and $\Omega\geq0$, which corresponds to the antiferromagnetic XXZ chain.\\
 The generation of entanglement, is a competition between the effects of the environment and the coupling between the two qubits. On one hand we have the case in which there is no couplings between the two qubits. Here the entanglement is generated via the interaction of the two qubits with a common environment. Such environmentally induced entanglement has been reported in \cite{31}.  Hence the environment, which is known to cause decoherence can never the less generate some entanglement between the two qubits.  A similar conclusion has been made by the authors of \cite{30}. Such environment induced entanglement is very fragile and increases with increasing coupling strength between the system and the bath.

 On the other hand we also have the case of entanglement generated through the coupling between the two qubits. The coupling generated entanglement is stronger and dominates the preceding case. For coupling induced entanglement, increasing the coupling strength between the two qubit systems and the environment will rather cause faster decay of entanglement.
To study the effects of the DM interaction on the entanglement dynamics, we consider an initially disentangled state . For this case of two initially separable qubits which become entangled in time through the effects of the environment, their entanglement is destroyed by increasing the strength of the DM interactions as seen in figure.1. Thus the DM interactions destroy environment induced entanglement but enhance coupling induced entanglement.
\begin{figure}
\resizebox{0.5\textwidth}{!}{%
  \includegraphics{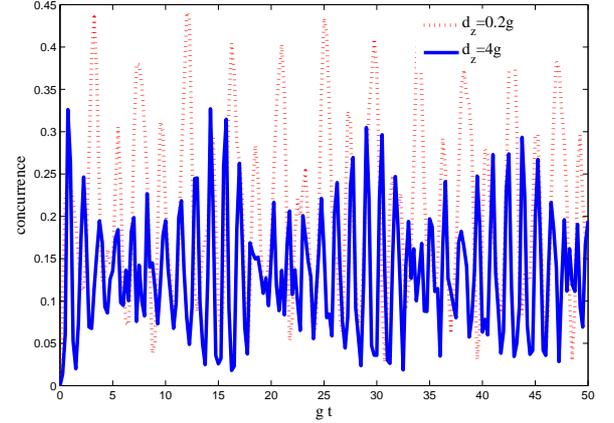}
}

\caption{Concurrence VS time for an initially disentangled qubit (i.e, $|\psi\rangle=|00\rangle$ ), for different values of $d_{z}$, with the corresponding values of the parameters $\Omega=0$, $\Gamma_{z}=0$, $g=g_{0}=2$, $\mu_{0}=2g$ et $T=2g$ .}\label{Fig:1}
\end{figure}
The behavior of an initially entangled spin chain is very different from that of a spin chain with no initial entanglement. We study the behavior of the concurrence for a maximally entangled initial qubit state $|\psi\rangle=\frac{1}{2}(|00\rangle+|11\rangle)$. We find that the effects of the DM interaction depend largely on the temperature, and on the Heisenberg couplings $\Gamma_{z}$ and $\Omega$. In the absence of the couplings i.e. $\Gamma_{z}=\Omega=0$ as seen in figure.2 the DM interactions preserve the entanglement and also greatly enhance the revival of entanglement. When the Heisenberg couplings set in this enhancement effect are reduced, the concurrence shows a sinusoidal oscillation in space as presented in figure.3. This is due to competing effects between the anti-symmetric DM interaction and the symmetric Heisenberg interactions. This is because in contrast to the Heisenberg interactions which tend to render neighbor spins parallel, the DM interaction has the effect of turning them perpendicular to one another.
\begin{figure}
\resizebox{0.5\textwidth}{!}{%
  \includegraphics{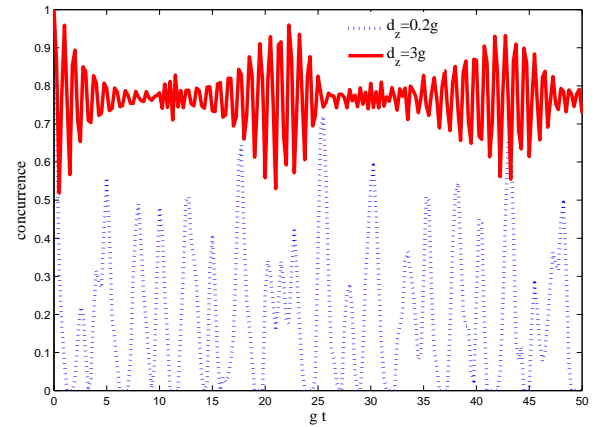}
}

\caption{Concurrence VS time for an initially maximally entangled state  $|\psi\rangle=\frac{1}{2}(|00\rangle+|11\rangle)$, for different values of $d_{z}$, with the corresponding values of the parameters $\Omega=0$, $\Gamma_{z}=0$, $g=g_{0}=2$, $\mu_{0}=2g$ and $T=3g$ .}\label{Fig:2}
\end{figure}

\begin{figure}
\resizebox{0.5\textwidth}{!}{%
  \includegraphics{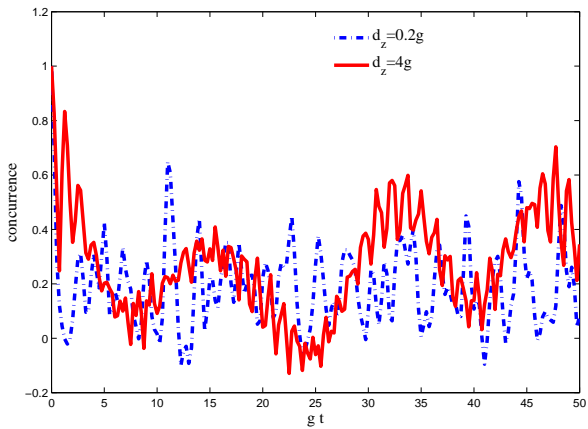}
}

\caption{Concurrence VS time for an initially maximally entangled state  $|\psi\rangle=\frac{1}{2}(|00\rangle+|11\rangle)$,for different values of $d_{z}$, with the corresponding values of the parameters $\Omega=1$, $\Gamma_{z}=0.5$, $g=g_{0}=2$, $\mu_{0}=2g$ et $T=2g$ .}\label{Fig:3}
\end{figure}
It is observed in figure.4 that, the entanglement decays more rapidly as the temperature increases. Increasing the temperature introduces thermal fluctuations which destroy quantum correlations. At low temperatures the entanglement exhibits periodic oscillations. At high temperatures we also observe the interesting phenomenon of entanglement sudden death (ESD) i.e. when the entanglement of the system is observed to suddenly disappear without any exponential decay. It was shown in \cite{32} that noisy environments may drive entanglement to vanish completely in finite time and they call the phenomenon entanglement sudden death. The sudden death of entanglement is a very undesirable effect since major quantum protocols for quantum computing; depend on the preservation of entanglement in the system. Here the non-Markovian environment is seen to cause the revival or rebirth of entanglement after ESD.
\begin{figure}
\resizebox{0.5\textwidth}{!}{%
  \includegraphics{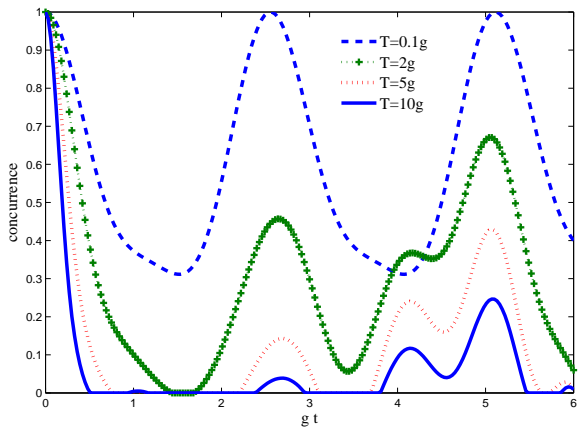}
}

\caption{Concurrence VS time for an initially maximally entangled state $|\psi\rangle=\frac{1}{2}(|00\rangle+|11\rangle)$, for different temperatures with the corresponding values of the parameters $\Omega=0$, $\Gamma_{z}=0$, $g=g_{0}=2$, $\mu_{0}=2g$ and $d_{z}=0.2g$ .}\label{Fig:4}
\end{figure}
The oscillatory collapse and revival behavior of the entanglement due to the influence of the non-Markovian environment can be understood by analogy to the collapse and revival of atomic population inversion of a two-level atom interacting with a single mode field in quantum optics. It is known that for a two level system coupled to an oscillating driving field the probability of being in the ground or excited states, exhibit oscillatory behavior (Rabi oscillations). Similarly in this study, the qubits are coupled to a single mode bath thus the quantum fluctuations of the system due to the bath may become uncorrelated in time leading to the collapse of entanglement. As time goes on these quantum fluctuations may again become correlated leading to the revival of entanglement. The DM interaction increases the frequency of the quantum fluctuations \cite{33} thus enhancing the entanglement. Furthermore this sudden death of entanglement can be avoided by increasing the strength of DM interaction as observed in figure.5.
\begin{figure}
\resizebox{0.5\textwidth}{!}{%
  \includegraphics{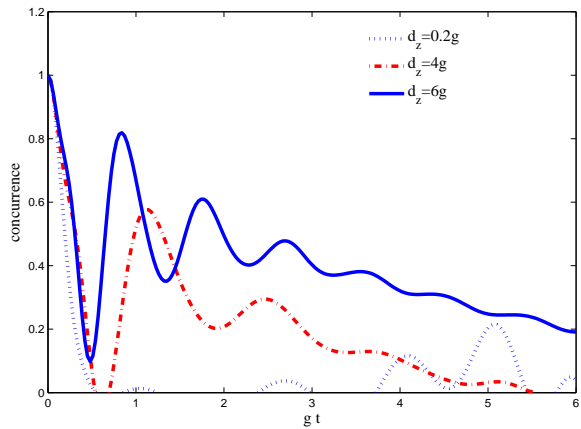}
}

\caption{Concurrence VS time for an initially maximally entangled state $|\psi\rangle=\frac{1}{2}(|00\rangle+|11\rangle)$ at high temperature (i.e, $T=10g$) for different values of $d_{z}$, with the corresponding values of the parameters $\Omega=0$, $\Gamma_{z}=0$, $g=g_{0}=2$, $\mu_{0}=2g$ .}\label{Fig:5}
\end{figure}

When coupling between the two qubits is switched on, the entanglement is observed to be preserved for a longer time. The dependence of the concurrence on the coupling strengths $\Gamma_{z}$ and $\Omega$ is closely linked. For example, we note that if $\Gamma_{z}>0$ then increasing $\Omega$  will cause the concurrence to reduce while if $\Gamma_{z}<0$ then increasing $\Omega$ will improve the concurrence. The same holds for the variation of $\Gamma_{z}$ with a fixed value of $\Omega$. In our numerical analysis, we note that the effective Heisenberg coupling between the two qubits can be written as $\chi=|\Gamma_{z}-\Omega|$. 		
The entanglement is enhanced by increasing the factor $\chi$ as seen in figure.6. The entanglement does not depend on how large $\Gamma_{z}$  and $\Omega$ are but rather depends on their difference. We find that for $\Gamma_{z}=\Omega$ , i.e. $\chi=0$  the concurrence is low and it increases as the value of $\chi$ increases.  From this we can conclude that the anisotropic XXZ chain will be better than the Isotropic XXX chain for preserving entanglement and hence for various quantum information processing tasks.

We also find that the entanglement decays very slowly when the system bath coupling is small and faster for a strong coupling between the system and the bath (see figure.7). This is so because in the case where the system, is strongly coupled to the bath, decoherence from the bath is more prominent and leads to faster decay of the quantum correlations.
When the coupling amongst the bath spins is strong we observe that the entanglement decays more slowly (such presented in figure.8).This is an indication that strong coupling amongst the bath spins effectively decouples the bath from the system thus preserving entanglement . As expected strong coupling between the two qubits also allows them to keep their entanglement for longer times.\\
\begin{figure}
\resizebox{0.5\textwidth}{!}{%
  \includegraphics{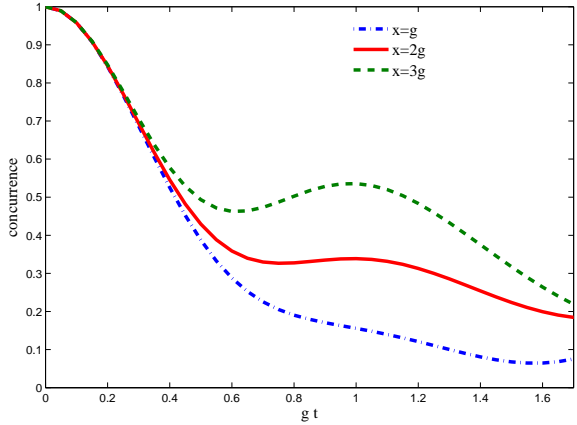}
}

\caption{Concurrence VS time for an initially maximally entangled state $|\psi\rangle=\frac{1}{2}(|00\rangle+|11\rangle)$  for different values of $\chi$ considering $\Omega=0$,with the corresponding values of the parameters $d_{z}=0$, $g=g_{0}=2$, $\mu_{0}=2g$, $T=3g$ .}\label{Fig:6}
\end{figure}
\begin{figure}
\resizebox{0.5\textwidth}{!}{%
  \includegraphics{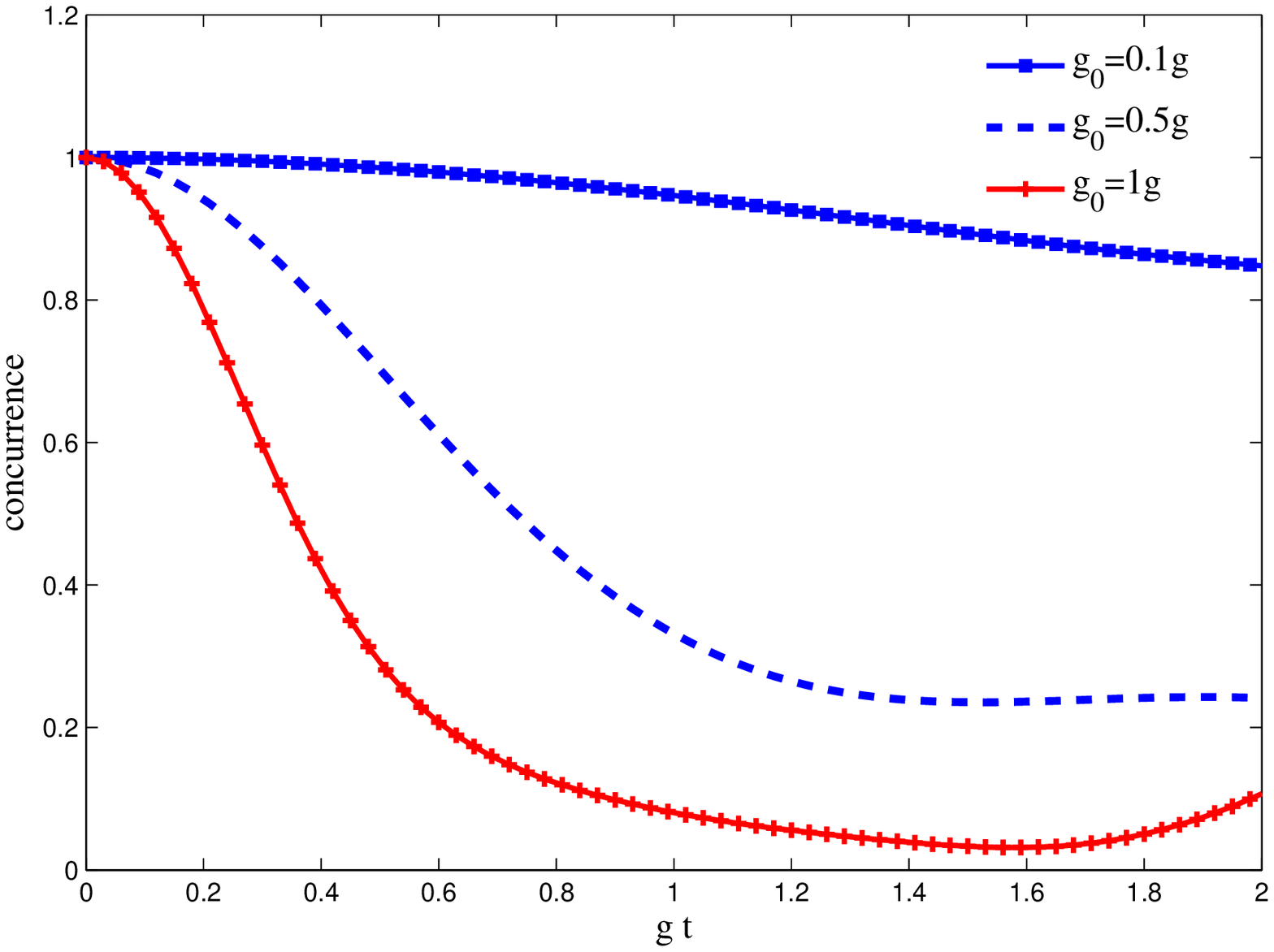}
}

\caption{Concurrence VS time for an initially maximally entangled state $|\psi\rangle=\frac{1}{2}(|00\rangle+|11\rangle)$  for different values of $g_{0}$, with the corresponding values of the parameters $\Omega=2$, $d_{z}=1$, $\Gamma_{z}=1$, $g=2$, $\mu_{0}=2g$, $T=2g$ .}\label{Fig:7}
\end{figure}
\begin{figure}
\resizebox{0.5\textwidth}{!}{%
  \includegraphics{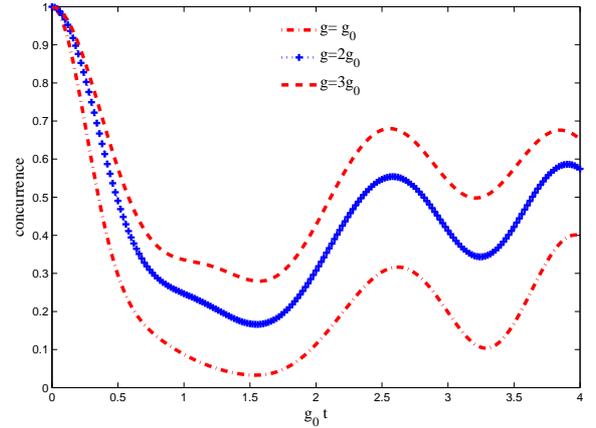}
}

\caption{Concurrence VS time for an initially maximally entangled state $|\psi\rangle=\frac{1}{2}(|00\rangle+|11\rangle)$  for different values of $g$, with the corresponding values of the parameters $\Omega=2$, $d_{z}=0.2$, $\Gamma_{z}=2$, $g_{0}=2$, $\mu_{0}=2g$, $T=3g_{0}$ .}\label{Fig:8}
\end{figure}
\section{discussion and concluding remarks}
\label{sec:4}
The entanglement dynamics for a system of two qubits XXZ spin chian coupled to antiferromagnetic  spin bath with DM interactions have been studied under the influence of an external magnetic field. The results obtained shows consequently the strong dependence with different coupling spin and on the nature of the bath.

Numerical analysis of the behavior of the concurrence vis-a-viz the various system parameters revealed that the effects of the DM interaction depend on the initial state of the system and on how the entanglement is generated. For an initially disentangled qubit pair, and in the absence of any coupling between the two qubits, the common bath can generate some effective entanglement between the two qubits. The DM interaction destroys such environmentally generated entanglement; this is contrary to the case of an initially entangled qubit where the DM interactions rather enhance the entanglement.

For the long time behavior of the entanglement for an initially entangled state, it was also observed to initially decay exponentially with time and then to undergo continues cycles of collapse and revival. This collapse and revival behavior is attributed to the non-Markovian nature of the bath in which memory effects of the bath can reconstruct the entanglement of the system, with time. It is seen that the DM interactions play an important role in the weak inter-qubit coupling limit, and for high temperatures where they delay the decay of entanglement and enhance revival oscillations in the entanglement.

Increasing the temperature can lead to appearance of the Entanglement sudden death effect. However this sudden death of entanglement can be avoided by increasing the strength of the DM interaction. The effects of $\Gamma_{z}$ and $\Omega$ on the entanglement are closely linked and we find that the effective Heisenberg coupling in the XXZ chain is given by $\chi=|\Gamma_{z}-\Omega|$.

Increasing  $\chi$ enhances and preserves entanglement while we see that entanglement decays faster for small $\chi$ even if  $\Gamma_{z}$ and $\Omega$  are both large. This suggests that the anisotropy XXZ spin chain is better than the isotropic XXX one(where $\chi=0$)  for QIP tasks. Further more strong intrabath coupling is seen to effectively decouple the bath from the system thus delaying the loss of entanglement in the system. Our result reveals that entanglement can be better preserved even for large finite temperatures and for a long time, by turning the strength of the external magnetic field, the DM interaction and the system parameters $\Gamma_{z}$ , $\Omega$ , $g_{0}$, $g$. We expect that the analysis done in this study will shed some light on the study of the dynamics of a multipartite entangled state under local noise. An interesting feature of this model is that it can be used as better quantum channel when entanglement transfer is considered. Therefore, in principle, it can be exploited as a quantum channel for teleportation with nonclassical fidelity at finite temperature, both very low and moderately low. That is, considering future research along these lines of investigation, it will be interesting to consider practical schemes for the realization of this kind of spin chains in highly controllable situations.

\end{document}